\documentclass[12pt,a4paper]{article}
\usepackage{amsmath,amsthm,amssymb, mathrsfs}
\usepackage{hyperref}
\usepackage{tikz}
\usepackage{graphicx}
\usepackage{cite}
\input xy
\xyoption{all}
\usepackage{epsfig}

\newtheorem{thm}{Theorem}[section]
\newtheorem{cor}[thm]{Corollary}
\newtheorem{lem}[thm]{Lemma}

\newtheorem{prop}[thm]{Proposition}

 \theoremstyle{definition}

\theoremstyle{remark}
\newtheorem{remark}[thm]{Remark}

\newtheorem{example}[thm]{Example}

\numberwithin{equation}{section}

\newcommand{\ben}{\begin{enumerate}}

\newcommand{\een}{\end{enumerate}}
\newcommand{\bit}{\begin{itemize}}
\newcommand{\eit}{\end{itemize}}

\begin{document}

\title
{Prony Scenarios and Error Amplification in a Noisy Spike-Train Reconstruction}

\author{Gil Goldman$^{1}$, Yehonatan Salman$^{2}$\\ and Yosef Yomdin$^{3}$
\\
\\
$^{1,2,3}$Department of Mathematics, The Weizmann Institute of\\
Science, Rehovot 76100, Israel\\
$^{1}$Email: gilgoldm@gmail.com\\
$^{2}$Email: salman.yehonatan@gmail.com\\
$^{3}$Email: yosef.yomdin@weizmann.ac.il}

\maketitle

\vskip-1cm\begin{abstract}

The paper is devoted to the characterization of the geometry of Prony curves arising from spike-train signals. We give a sufficient condition which guarantees the blowing up of the amplitudes of a Prony curve $S$ in case where some of its nodes tend to collide. We also give sufficient conditions on $S$ which guarantee a certain asymptotic behavior of its nodes near infinity.

\end{abstract}

\section{Introduction and Mathematical Background}

The main object of study in this paper are spike-train signals $F$ of the form
\begin{equation}\hskip-6.7cm F(x) = \sum_{i = 1}^{d}a_{i}\delta(x - x_{i}), a_{i}, x_{i}\in\Bbb R\end{equation}
assuming that the number of summands $d$ is known while the values of the amplitudes $A = (a_{1},...,a_{d})$ and the nodes $X = (x_{1},...,x_{d})$ are to be obtained.

Our aim is to find the best approximation of the parameters $a_{1},...,a_{d},x_{1},...,x_{d}$ given the following $q + 1$ moments
\begin{equation}\hskip-5cm m_{k}(F) = \int_{-\infty}^{\infty}x^{k}F(x)dx = \sum_{i = 1}^{d}a_{i}x_{i}^{k}, k = 0,...,q\end{equation}
of the signal $F$ where $0\leq q\leq 2d - 1$. The system (1.2) is called a Prony system of equations for the signal $F$. In case where $q = 2d - 1$ we will call the above system a \textbf{complete Prony system} for the parameters $a_{1},...,a_{d},x_{1},...,x_{d}$.

More generally, for each fixed point $\mu = (\mu_{0},...,\mu_{2d - 1})\in\Bbb R^{2d}$ and $0\leq q\leq 2d - 1$ the following system
\vskip-0.2cm
$$\hskip-9.5cm a_{1} + a_{2} + ... + a_{d} = \mu_{0},$$
$$\hskip-8.25cm a_{1}x_{1} + a_{2}x_{2} + ... + a_{d}x_{d} = \mu_{1},$$
\begin{equation}\hskip-11cm...........\end{equation}
$$\hskip-8.25cm a_{1}x_{1}^{q} + a_{2}x_{2}^{q} + ... + a_{d}x_{d}^{q} = \mu_{q},$$
defines (in the general case) a $2d - q - 1$ dimensional algebraic variety $S_{q}(\mu)$ in $\Bbb R^{2d}$ where each point $(A,X)\in S_{q}(\mu)$ can be identified with a signal $F$ whose set of amplitudes and nodes are given respectively by the points $A$ and $X$ in $\Bbb R^{d}$.

In this paper we will be dealing with the special case where $q = 2d - 2$ for which $S_{2d - 2}(\mu)$ defines, in the general case, a curve in $\Bbb R^{2d}$. For this case we assume that $\mu\in\Bbb R^{2d - 1}$ and define its corresponding Hankel matrix $\mathcal{M} = \mathcal{M}(\mu)$ by
\vskip-0.2cm
$$\hskip-7.5cm \mathcal{M} = \left(\begin{array}{cccc}
\mu_{0} & \mu_{1} & ... & \mu_{d - 1}\\
\mu_{1} & \mu_{2} & ... & \mu_{d}\\
.........\\
\mu_{d - 1} & \mu_{d} & ... & \mu_{2d - 2}\\
\end{array}\right).$$
Observe that if each component of $\mathcal{M}$ is considered as a function of $A$ and $X$, where we replace each value $\mu_{i}$, in the definition of $\mathcal{M}$ above, with its left hand side in the system of equations (1.3), then $\mathcal{M}$ is constant on $S_{2d - 2}(\mu)$.

Given a Prony system of equations (1.3), where $q = 2d - 2$, we will all ways assume that the Hankel matrix $\mathcal{M}$, corresponding to the point $\mu$, is non degenerate. This assumption is necessary since, as we will show later (see Remark 1.3), the degeneracy of the matrix $\mathcal{M}$ is equivalent to the assumption that at least two nodes $x_{i}, x_{j}$ where $i\neq j$ collide identically or that at least one amplitude $a_{i}$ vanishes identically on $S_{2d - 2}(\mu)$. However, a signal $F$ of the form (1.1) cannot be contained on such curves since this will imply that the number of summands of the signal $F$ is less than $d$.\\

Prony's type systems of equations have been investigated by many authors (see \cite{1, 2, 3, 4, 5, 6, 7, 9, 22, 24, 25, 26}) and have been found out to be useful in many fields in applied mathematics such as in imaging in the context of superresolution (\hskip-0.001cm\cite{8, 10, 12, 13, 14, 15, 16, 17, 18, 19, 20, 21}) and in approximation theory (\hskip-0.001cm\cite{11, 23}). Our motivation in this paper comes from an important result, found in \cite{1}, about the worst case errors, in the presence of noise, in the reconstruction of a solution to the complete Prony system and the reconstruction of the Prony curve $S_{2d - 2}$.

More specifically, in \cite{1} the authors prove that in case where the nodes $x_{1},...,x_{d}$ form a cluster of size $h\ll1$, while the measurements error of the acquired moments (1.2) in the presence of noise is of order $\epsilon$, the worst case error in the reconstruction of the solution to the complete Prony system is of order $\epsilon h^{-2d + 1}$. However, it is also proved in \cite{1} that the worst case error in the reconstruction of the Prony curve $S_{2d - 2}$ is of order $\epsilon h^{-2d + 2}$. That is, the reconstruction of the Prony curve $S_{2d - 2}$ is $h$ times better than the reconstruction of the solutions themselves and this curve provides a rather accurate prediction of the possible behavior of the noisy reconstructions.

In other words, from the noisy measurements $\mu_{0},...,\mu_{2d - 1}$ we can first reconstruct the Prony curve $S_{2d - 2}(\mu)$ with the improved accuracy, and then we can consider the curve $S_{2d - 2}(\mu)$ as a prediction of a possible distribution of the noisy reconstructions. This scenario includes an accurate description of the behavior of the nodes $x_{j}$ and the amplitudes $a_{j}$ along the curve $S_{2d - 2}(\mu)$.

Hence, since the incorrect reconstructions, caused by the measurements noise, are spread along the Prony curve $S_{2d - 2}$, our main aim in this paper is to investigate the behavior of the amplitudes on $S_{2d - 2}$ in case where some of its nodes tend to collide. We will also investigate the asymptotic behavior of the nodes on $S_{2d - 2}$ at infinity.

The main results of this paper are given in Theorems 2.1 and 2.2 in Sect. 2 below. In Theorem 2.1 we prove that in the general case any collision of two nodes $x_{i}$ and $x_{j}$ where $i\neq j$ on a Prony curve $S_{2d - 2}(\mu)$ results in the blowing up of their corresponding amplitudes $a_{i}$ and $a_{j}$, that is $|a_{i}|, |a_{j}|\rightarrow\infty$. In Theorem 2.2 we characterize the asymptotic behavior near infinity of the nodes on the Prony curve $S_{2d - 2}(\mu)$ and show that in the general case all the nodes are bounded in absolute value except maybe only one node. In Sect. 3 we prove our main results, Theorems 1.1 and 1.2, and in Sect. 4 we investigate the geometry of $S_{2d - 2}(\mu)$ for the special cases where $d = 2, 3$.\\

Before formulating the main results we will introduce some basic notations, definitions and results concerning Prony curves. Denote by $\mathcal{P}_{d}^{A}$ and $\mathcal{P}_{d}^{X}$ the following sets
\vskip-0.2cm
$$\hskip-8.25cm\mathcal{P}_{d}^{A} = \{A = (a_{1},a_{2},...,a_{d})\in\Bbb R^{d}\}$$
$$\hskip-4.55cm\mathcal{P}_{d}^{X} = \{X = (x_{1},x_{2},...,x_{d})\in\Bbb R^{d}:x_{1} < x_{2} < ... < x_{d}\}$$
which we call the amplitudes parameter space and nodes parameter space respectively. The restriction that $x_{i} < x_{j}$ for $i < j$ is imposed since any permutation of the nodes, and their corresponding amplitudes, of a solution to the Prony system of equations (1.3) is also a solution. Thus, we omit redundant data by taking only one permutation of each set of solutions to (1.3). With this notation we define the parameter space $\mathcal{P}_{d}$ of signals by $\mathcal{P}_{d} = \mathcal{P}_{d}^{A}\times\mathcal{P}_{d}^{X}$, i.e., every signal $F$ given by (1.1) is identified with the point $(A, X)\in\mathcal{P}_{d}$ where $A$ and $X$ denote respectively the amplitudes and nodes vectors of $F$. We also assume that each Prony curve $S_{2d - 2}(\mu)$ is contained in $\mathcal{P}_{d}$. We denote by $S_{2d - 2}^{X}(\mu)$ the projection of $S_{2d - 2}(\mu)$ to the nodes parameter space $\mathcal{P}_{d}^{X}$.\\

\hskip-0.6cm For every $1\leq i\leq d$ define the projection $\pi_{i}:\Bbb R^{d}\rightarrow\Bbb R^{d - 1}$ by
$$\hskip-6.5cm \pi_{i}(x_{1},...x_{d}) = (x_{1},...,x_{i - 1},x_{i + 1},...,x_{d}).$$
Define the following symmetric polynomials $\varrho_{k}, k = 1,...,d - 1$ and $\sigma_{k}, k = 1,...,d$ in $\Bbb R^{d - 1}$ and $\Bbb R^{d}$ respectively by :
\vskip-0.2cm
$$\hskip-2.4cm\varrho_{k}\left(x_{1},...,x_{d - 1}\right) = (- 1)^{k}\sum_{1\leq i_{1} < ... < i_{k}\leq d - 1}x_{i_{1}}...x_{i_{k}}, 1\leq k\leq d - 1,$$
$$\hskip-3.5cm\sigma_{k}(x_{1},...,x_{d}) = (- 1)^{k}\sum_{1\leq i_{1} < ... < i_{k}\leq d}x_{i_{1}}...x_{i_{k}}, 1\leq k\leq d.$$
Observe that the symmetric polynomials $\sigma_{1},...,\sigma_{d}$ satisfy, for each $z\in\Bbb C$, the following equation
\begin{equation}Q(z) = (z - x_{1})(z - x_{2})...(z - x_{d}) = z^{d} + \sigma_{1}(X)z^{d - 1} + ... + \sigma_{d - 1}(X)z + \sigma_{d}(X)\end{equation}
and can be in fact equivalently defined as the unique functions which satisfy equation (1.4) for each complex number $z$. We identify the set of monic polynomials of order $d$ with the space $\mathcal{W}\cong\Bbb R^{d}$ where the ordered set $\sigma = (\sigma_{1},...,\sigma_{d})$ of the coefficients of the polynomial $Q$ is identified with $Q$.

By equation (1.4) each fixed point $X$ in $\mathcal{P}_{d}^{X}$ corresponds to a unique set of coefficients $\sigma_{1},...,\sigma_{d}$ and thus to a point $\sigma$ in $\mathcal{W}$. However, not every point $\sigma$ in $\mathcal{W}$ corresponds to a point in $\mathcal{P}_{d}^{X}$ since, as can be seen from equation (1.4), there is no guarantee that all the roots of the polynomial $Q$ are real and distinct if its coefficients are chosen arbitrarily. Hence, we denote by $H_{d}\subset\mathcal{W}$ the subset of hyperbolic polynomials of all points whose components from left to right are the coefficients $\sigma_{1},...,\sigma_{d}$ of a polynomial $Q$, given as in equation (1.4), whose roots are real and distinct. With this notation we define the Vieta mapping
\vskip-0.2cm
$$\hskip-11.7cm\mathcal{V}_{d}:\mathcal{P}_{d}^{X}\rightarrow H_{d}$$
$$\hskip-6cm\mathcal{V}_{d}\left(X\right) = \left(\sigma_{1}\left(X\right),...,\sigma_{d}\left(X\right)\right), X = (x_{1},...,x_{d})$$
which is a bijection between $\mathcal{P}_{d}^{X}$ and $H_{d}$.

We will now formulate the following important lemma which is used during the text and its proof is given in the Appendix.

\begin{lem}

For $d\leq q\leq 2d - 2$ the projection $S_{q}^{X}(\mu)$ of the surface $S_{q}(\mu)$, defined by the system of equations (1.3), into the nodes parameter space $\mathcal{P}_{d}^{X}$ is given by the following system of equations

$$\hskip-5.15cm \mu_{0}\sigma_{d}\left(X\right) + \mu_{1}\sigma_{d - 1}\left(X\right) + ... + \mu_{d - 1}\sigma_{1}\left(X\right) + \mu_{d} = 0,$$
$$\hskip-5.15cm \mu_{1}\sigma_{d}\left(X\right) + \mu_{2}\sigma_{d - 1}\left(X\right) + ... + \mu_{d}\sigma_{1}\left(X\right) + \mu_{d + 1} = 0,$$
\begin{equation}\hskip-7cm..............\end{equation}
$$\hskip-4.05cm \mu_{q - d}\sigma_{d}\left(X\right) + \mu_{q - d + 1}\sigma_{d - 1}\left(X\right) + ... + \mu_{q - 1}\sigma_{1}\left(X\right) + \mu_{q} = 0.$$
\end{lem}

\vskip0.3cm

From Lemma 1.1 it follows that the non degeneracy condition on $\mathcal{M}$ implies that $S_{2d - 2}(\mu)$ is a curve in $\mathcal{P}_{d}$. Indeed, by Lemma 1.1 the projection $S_{2d - 2}^{X}(\mu)$ to the nodes parameter space $\mathcal{P}_{d}^{X}$ is given by the following system of equations

\begin{equation}\hskip-1.3cm \mu_{k}\sigma_{d}\left(X\right) + \mu_{k + 1}\sigma_{d - 1}\left(X\right) + ... + \mu_{k + d - 1}\sigma_{1}\left(X\right) = - \mu_{d + k}, k = 0,...,d - 2.\end{equation}
\vskip0.3cm
\hskip-0.6cm Now we "complete" the system (1.6) by adding the following equation
\begin{equation}\hskip-5.5cm \mu_{d - 1}\sigma_{d}\left(X\right) + \mu_{d}\sigma_{d - 1}\left(X\right) + ... + \mu_{2d - 2}\sigma_{1}\left(X\right) = t\end{equation}
where $t$ is a real parameter. Since $\det\mathcal{M}\neq0$ it follows that the system of equations obtained by combining the system (1.6) and equation (1.7) is non degenerate. Hence, each one of the variables $\sigma_{1},...,\sigma_{d}$ can be expressed via Cramer's rule as a linear function of $t$. Explicitly we have
\begin{equation}\hskip-5.85cm\sigma_{d - k + 1} = \frac{(- 1)^{d + k}\mathcal{M}_{d,k}}{\det \mathcal{M}}t + b_{d - k + 1}, k = 1,...,d,\end{equation}
where
\begin{equation}\hskip-1.1cm b_{d - k + 1} = \frac{(- 1)^{k}}{\det\mathcal{M}}\left(\mu_{d}\cdot\mathcal{M}_{1, k} - \mu_{d + 1}\cdot\mathcal{M}_{2, k} + ... + (- 1)^{d - 2}\mu_{2d - 2}\cdot\mathcal{M}_{d - 1,k}\right)\end{equation}
and where $\mathcal{M}_{i,j}$ denotes the minor of the entry in the $i$-th row and $j$-th column of $\mathcal{M}$. Since the coefficients $\sigma_{1},...,\sigma_{d}$ correspond to $d$ real nodes $x_{1},...,x_{d}$ we take only those values of $t$ for which the line $l_{\mu}$, defined by the parametrization (1.8), is in $H_{d}$. We define $A_{\mu}\subseteq\Bbb R$ as the set of all $t\in\Bbb R$ for which $l_{\mu}(t)\subset H_{d}$. It can be easily seen that $A_{\mu}$ is a finite union of open intervals in $\Bbb R$.

Since the Vieta mapping $\mathcal{V}_{d}$ is a bijection between $H_{d}$ and the nodes parameter space $\mathcal{P}_{d}^{X}$ it follows that each node $x_{i}$ can also be parameterized as a function of $t$ where $t\in A_{\mu}$. For the amplitudes $a_{1},...,a_{d}$ we can solve the first $d$ equations in the system (1.3) which is of Vandermonde's type if $x\in \mathcal{P}_{d}^{X}$ and express each one of these amplitudes as a function of $x_{1},...,x_{d}$ (and thus as a function of $t$). Hence, we obtained the following proposition.

\begin{prop}

Let $\mu$ be a point in $\Bbb R^{2d - 1}$ and assume that its Hankel matrix $\mathcal{M}(\mu)$ is non degenerate. Then, the set of solutions $S_{2d - 2}(\mu)$ in $\mathcal{P}_{d}$ to the system of equations (1.3) for $q = 2d - 2$ is a curve whose projection $S_{2d - 2}^{X}(\mu)$ to $\mathcal{P}_{d}^{X}$ can be parameterized by $\mathcal{V}_{d}^{-1}(\sigma(t)), t\in A_{\mu}$ where $\sigma(t) = (\sigma_{1}(t),...,\sigma_{d}(t))$ is given by equations (1.8)-(1.9).

\end{prop}

\begin{remark}

The assumption that the Hankel matrix $\mathcal{M}$ is non degenerate is equivalent to the assumption that no two nodes $x_{i}$ and $x_{j}$ collide identically and that no amplitude $a_{i}$ vanishes identically on $S_{2d - 2}(\mu)$. Indeed, $S_{2d - 2}(\mu)$ is defined by the system of equations (1.3), where $q = 2d - 2$, and this system can be equivalently written in the following matrix form $\mathcal{M} = V\Lambda V^{T}$ where
\begin{equation}\hskip-2.5cm V = \left(\begin{array}{cccc}
1 & 1 & ... & 1\\
x_{1} & x_{2} & ... & x_{d}\\
& .........\\
x_{1}^{d - 1} & x_{2}^{d - 1} & ... & x_{d}^{d - 1}\\
\end{array}\right),
\Lambda = \left(\begin{array}{cccc}
a_{1} & 0 & ... & 0\\
0 & a_{2} & ... & 0\\
& ..........\\
0 & 0 & ... & a_{d}\\
\end{array}\right).\end{equation}
\vskip0.1cm
Hence, the degeneracy of $\mathcal{M}$ is equivalent to the degeneracy of the Vandermonde matrix $V$ or of the diagonal matrix $\Lambda$. The degeneracy of $V$ is equivalent to the condition that at least two nodes $x_{i}$ and $x_{j}$, where $i\neq j$, coincide while the degeneracy of $\Lambda$ is equivalent to the condition that at least one of the amplitudes $a_{i}$ vanishes.

\end{remark}

\begin{remark}

Observe that since $S_{2d - 2}(\mu)\subset\mathcal{P}_{d}$ it follows in particular that for each point $(A,X)\in S_{2d - 2}(\mu)$ we have $X\in\mathcal{P}_{d}^{X}$ and thus collision of two nodes $x_{i}$ and $x_{i + 1}$ cannot actually occur on $S_{2d - 2}(\mu)$ since $x_{i} < x_{i + 1}$. However, when we say that two nodes collide on $S_{2d - 2}$ at $t = t_{0}$ and write $x_{i} \sim x_{i + 1}$ we mean that $x_{i + 1}(t) - x_{i}(t)\rightarrow0$ as $t\rightarrow t_{0}$ where $t_{0}$ is a boundary point of $A_{\mu}$.

\end{remark}

\section{Main Results}

\hskip0.6cm The first main result, Theorem 2.1, implies that in the general case any collision of two nodes on a Prony curve $S_{2d - 2}$ results in the blowing up of their corresponding amplitudes. The exact formulation is as follows:

\begin{thm}

Let $\mu$ be a point in $\Bbb R^{2d - 1}$ and assume that its corresponding Hankel matrix $\mathcal{M}$ is non degenerate. Then, if the nodes $x_{i}(t)$ and $x_{i + 1}(t)$ tend to collide on $S_{2d - 2}(\mu)$ as $t\rightarrow t_{0}$, where $1\leq i\leq d - 1$, then the amplitudes $a_{i}(t)$ and $a_{i + 1}(t)$ tend to infinity as $t\rightarrow t_{0}$.
\end{thm}

The second main result, Theorem 2.2, implies that in the general case only one node on a Prony curve $S_{2d - 2}$ can approach to infinity at a time. The exact formulation is as follows:

\begin{thm}

Assume that the Hankel matrix $\mathcal{M}$, of a point $\mu\in\Bbb R^{2d - 1}$, and its top-left $(d - 1)\times(d - 1)$ sub matrix are non degenerate. Let $A_{\mu}'\subset A_{\mu}$ be an unbounded interval of $A_{\mu}$ and let
\vskip-0.2cm
$$\hskip-6.5cm\gamma(t) = (x_{1}(t),x_{2}(t),...,x_{d}(t)), t\in A_{\mu}'$$
be the parametrization, of a connected component of $S_{2d - 2}^{X}$, given as in Proposition 1.2 which is now restricted on $A_{\mu}'$. That is, $\gamma(t) = \mathcal{V}_{d}^{-1}(\sigma(t)), t\in A_{\mu}'$ where $\sigma(t)$ is given by equations (1.8)-(1.9). Then, as $t\rightarrow\infty$ (or $t\rightarrow-\infty$) in $A_{\mu}'$ at most one node $x_{i} = x_{i}(t)$ tends to infinity.
\end{thm}

\section{Proofs of the Main Results}

\textbf{Proof of Theorem 2.1}: First observe that if two nodes tend to collide on the Prony curve $S_{2d - 2}(\mu)$ then it follows immediately, from the factorization $\mathcal{M} = V\Lambda V^{T}$ where $V$ and $\Lambda$ are given by (1.10), that at least one amplitude must tend to infinity. Indeed, if $t_{0}$ is a point for which $x_{i + 1}(t) - x_{i}(t)\rightarrow0$ as $t\rightarrow t_{0}$ then from the definition of the matrix $V$ we have that $\det V(t)\rightarrow0$. Hence, it follows that
\begin{equation}\hskip-3.25cm |a_{1}(t)|...|a_{d}(t)| = |\det\Lambda(t)| = \left|\frac{\det \mathcal{M}}{(\det V(t))^{2}}\right|\rightarrow\infty\hskip0.25cm\mathrm{as}\hskip0.25cm t\rightarrow t_{0}\end{equation}
since by assumption $\det\mathcal{M}\neq0$. Hence, from equation (3.1) there exists an amplitude $a_{j}$ satisfying $|a_{j}(t)|\rightarrow\infty$ as $t\rightarrow t_{0}$. However, from the above analysis it is still not clear which amplitudes must tend to infinity where our goal is to prove that these are the amplitudes $a_{i}$ and $a_{i + 1}$ corresponding to the nodes $x_{i}$ and $x_{i + 1}$.

For this we will have to express the amplitudes $a_{k}, k = 1,2,...,d$ in terms of $x_{i}, i = 1,2,...,d$. Observe that the first $d$ equations in the system (1.3) can be rewritten as follows

$$\hskip-4.5cm\left(\begin{array}{cccc}
1 & 1 & ... & 1\\
x_{1} & x_{2} & ... & x_{d}\\
........\\
x_{1}^{d - 1} & x_{2}^{d - 1} & ... & x_{d}^{d - 1}
\end{array}\right)
\left(\begin{array}{c}
a_{1}\\
a_{2}\\
...\\
a_{d}
\end{array}\right)
= \left(\begin{array}{c}
\mu_{0}\\
\mu_{1}\\
...\\
\mu_{d - 1}
\end{array}\right).$$
\vskip0.3cm
\hskip-0.6cm Using the well known formula for the inverse of the Vandermonde's matrix (see, for example \cite{27}) we obtain from the last matrix equation that
\begin{equation}\hskip-1.5cm\left(\begin{array}{c}
a_{1}\\
a_{2}\\
...\\
a_{d}
\end{array}\right)
 = \left(\begin{array}{ccccc}
\frac{\varrho_{d - 1}(\pi_{1}(X))}{L_{1}(X)} & \frac{\varrho_{d - 2}(\pi_{1}(X))}{L_{1}(X)} & ... & \frac{\varrho_{1}(\pi_{1}(X))}{L_{1}(X)} & \frac{1}{L_{1}(X)}\\
\frac{\varrho_{d - 1}(\pi_{2}(X))}{L_{2}(X)} & \frac{\varrho_{d - 2}(\pi_{2}(X))}{L_{2}(X)} & ... & \frac{\varrho_{1}(\pi_{2}(X))}{L_{2}(X)} & \frac{1}{L_{2}(X)}\\
..........\\
\frac{\varrho_{d - 1}(\pi_{d}(X))}{L_{d}(X)} & \frac{\varrho_{d - 2}(\pi_{d}(X))}{L_{d}(X)} & ... & \frac{\varrho_{1}(\pi_{d}(X))}{L_{d}(X)} & \frac{1}{L_{d}(X)}\\
\end{array}\right)
\left(\begin{array}{c}
\mu_{0}\\
\mu_{1}\\
...\\
\mu_{d - 1}
\end{array}\right)\end{equation}
where
$$\hskip-9cm L_{k}(X) = \underset{i\neq k}{\overset{d}{\underset{i = 1}{\prod}}}(x_{k} - x_{i}).$$
Let us assume without loss of generality that the nodes $x_{d - 1}$ and $x_{d}$ collide on $S_{2d - 2}(\mu)$, we will show that $a_{d - 1}$ and $a_{d}$ tend to infinity in this case. Also, from the symmetry of the formulas for $a_{d - 1}$ and $a_{d}$ in terms of the nodes $x_{1},x_{2},...,x_{d}$ it will be enough to prove our assertion only for $a_{d}$. Hence, we will concentrate from now on only on this amplitude.

For the amplitude $a_{d}$ we define the following polynomial
\vskip-0.2cm
$$\hskip-12cm P(x_{1},x_{2},...,x_{d - 1})$$
 $$ \hskip-1cm = \mu_{0}\varrho_{d - 1}(x_{1},...,x_{d - 1}) + \mu_{1}\varrho_{d - 2}(x_{1},...,x_{d - 1}) + ... + \mu_{d - 2}\varrho_{1}(x_{1},...,x_{d - 1}) + \mu_{d - 1}.$$
\vskip0.3cm
\hskip-0.6cm By equation (3.2) we have
\vskip0.15cm
$$\hskip-6.5cm a_{d}(x_{1},...,x_{d}) = \frac{P(x_{1},...,x_{d - 1})}{(x_{d} - x_{1})...(x_{d} - x_{d - 1})}$$
and thus if a point $(x_{1},x_{2},...,x_{d})\in S_{2d - 2}^{X}(\mu)$ satisfies that its nodes $x_{d - 1}$ and $x_{d}$ tend to collide then $a_{d}$ must tend to infinity unless the polynomial $P$ vanishes at this point (in which case $a_{d}$ may or may not tend to infinity). By Lemma 1.1 the curve $S_{2d - 2}^{X}(\mu)$ is defined by the following system of equations
$$\hskip-3.45cm \mu_{d - 1}(x_{1} + x_{2} + ... + x_{d}) - \mu_{d - 2}(x_{1}x_{2} + x_{1}x_{3} + ... + x_{d - 1}x_{d})$$ $$\hskip2cm + ... + ( - 1)^{d - 1}\mu_{0}x_{1}x_{2}...x_{d} = \mu_{d},$$
$$\hskip-3.8cm \mu_{d}(x_{1} + x_{2} + ... + x_{d}) - \mu_{d - 1}(x_{1}x_{2} + x_{1}x_{3} + ... + x_{d - 1}x_{d})$$ $$\hskip2.4cm + ... + ( - 1)^{d - 1}\mu_{1}x_{1}x_{2}...x_{d} = \mu_{d + 1},$$
$$\hskip-7cm...................$$
$$\hskip-3cm \mu_{2d - 3}(x_{1} + x_{2} + ... + x_{d}) - \mu_{2d - 4}(x_{1}x_{2} + x_{1}x_{3} + ... + x_{d - 1}x_{d})$$ $$\hskip3.1cm + ... + ( - 1)^{d - 1}\mu_{d - 2}x_{1}x_{2}...x_{d} = \mu_{2d - 2}.$$
\vskip0.3cm
\hskip-0.6cm By our assumption we have $x_{d} \sim x_{d - 1}$ and thus the last system of equations can be rewritten as
\vskip-0.5cm
$$\hskip-0.5cm \mu_{d - 1}(x_{d - 1} + x_{1} + ... + x_{d - 1}) - \mu_{d - 2}((x_{1} + ... + x_{d - 1})\cdot x_{d - 1} + x_{1}x_{2} + x_{1}x_{3}$$ $$\hskip-3.3cm  + ... + x_{d - 2}x_{d - 1}) + ... + ( - 1)^{d - 1}\mu_{0} (x_{1}x_{2}...x_{d - 1})\cdot x_{d - 1} = \mu_{d},$$
$$\hskip-0.7cm \mu_{d}(x_{d - 1} + x_{1} + ... + x_{d - 1}) - \mu_{d - 1}((x_{1} + ... + x_{d - 1})\cdot x_{d - 1} + x_{1}x_{2} + x_{1}x_{3}$$ $$\hskip-3.0cm  + ... + x_{d - 2}x_{d - 1}) + ... + ( - 1)^{d - 1}\mu_{1}(x_{1}x_{2}...x_{d - 1})\cdot x_{d - 1} = \mu_{d + 1},$$
\begin{equation}\hskip-8cm...................\end{equation}
$$\hskip-0.2cm \mu_{2d - 3}(x_{d - 1} + x_{1} + ... + x_{d - 1}) - \mu_{2d - 4}((x_{1} + ... + x_{d - 1})\cdot x_{d - 1} + x_{1}x_{2} + x_{1}x_{3} $$ $$\hskip-2.5cm  + ... + x_{d - 2}x_{d - 1}) + ... + ( - 1)^{d - 1}\mu_{d - 2}(x_{1}x_{2}...x_{d - 1})\cdot x_{d - 1} = \mu_{2d - 2}.$$
\vskip0.3cm
Hence, it will be enough to show that if a point $X^{\ast} = (x_{1},x_{2},...,x_{d - 1})$ in $\Bbb R^{d - 1}$ satisfies the system of equations (3.3) then $P(X^{\ast})\neq0$. Suppose that this is not the case, then there exists a point $X^{\ast}\in\Bbb R^{d - 1}$ which satisfies the system of equations (3.3) and is also a zero of the polynomial $P$. In terms of the symmetric polynomials  $\varrho_{1},...,\varrho_{d - 1}$ the last condition can be written as

$$\hskip-3.9cm \mu_{d - 1}(x_{d - 1} - \varrho_{1}(X^{\ast})) + \mu_{d - 2}(\varrho_{1}(X^{\ast})x_{d - 1} - \varrho_{2}(X^{\ast})) $$ $$ \hskip-0.6cm + ... + \mu_{1}(\varrho_{d - 2}(X^{\ast})x_{d - 1} - \varrho_{d - 1}(X^{\ast})) + \mu_{0}\varrho_{d - 1}(X^{\ast})x_{d - 1} = \mu_{d},$$
$$\hskip-4.2cm \mu_{d}(x_{d - 1} - \varrho_{1}(X^{\ast})) + \mu_{d - 1}(\varrho_{1}(X^{\ast})x_{d - 1} - \varrho_{2}(X^{\ast})) $$ $$ \hskip-0.2cm + ... + \mu_{2}(\varrho_{d - 2}(X^{\ast})x_{d - 1} - \varrho_{d - 1}(X^{\ast})) + \mu_{1}\varrho_{d - 1}(X^{\ast})x_{d - 1} = \mu_{d + 1},$$
\begin{equation}\hskip-6cm.............\end{equation}
$$\hskip-3.6cm \mu_{2d - 3}(x_{d - 1} - \varrho_{1}(X^{\ast})) + \mu_{2d - 4}(\varrho_{1}(X^{\ast})x_{d - 1} - \varrho_{2}(X^{\ast})) $$ $$ \hskip0.7cm + ... + \mu_{d - 1}(\varrho_{d - 2}(X^{\ast})x_{d - 1} - \varrho_{d - 1}(X^{\ast})) + \mu_{d - 2}\varrho_{d - 1}(X^{\ast})x_{d - 1} = \mu_{2d - 2},$$
$$\hskip-4cm - \mu_{d - 2}\varrho_{1}(X^{\ast}) - \mu_{d - 3}\varrho_{2}(X^{\ast}) - ... - \mu_{0}\varrho_{d - 1}(X^{\ast}) = \mu_{d - 1}.$$
\vskip0.3cm
\hskip-0.6cm From the last equation of the system (3.4) we have the following equality
$$\hskip-1.5cm\mu_{0}\varrho_{d - 1}(X^{\ast}) = - \mu_{1}\varrho_{d - 2}(X^{\ast}) - ... - \mu_{d - 3}\varrho_{2}(X^{\ast}) - \mu_{d - 2}\varrho_{1}(X^{\ast}) - \mu_{d - 1}.$$

\hskip-0.6cm Inserting this equality into the first equation of the system (3.4) we have
$$\mu_{d} = \mu_{d - 1}(x_{d - 1} - \varrho_{1}(X^{\ast})) + \mu_{d - 2}(\varrho_{1}(X^{\ast})x_{d - 1} - \varrho_{2}(X^{\ast})) + ... + \mu_{1}(\varrho_{d - 2}(X^{\ast})x_{d - 1} - \varrho_{d - 1}(X^{\ast}))$$ $$ \hskip-3cm + \left(-\mu_{1}\varrho_{d - 2}(X^{\ast}) - ... - \mu_{d - 3}\varrho_{2}(X^{\ast}) - \mu_{d - 2}\varrho_{1}(X^{\ast}) - \mu_{d - 1}\right)x_{d - 1}.$$
Observe that all the terms which contain a product of a symmetric polynomial with the node $x_{d - 1}$ cancel each other. Hence we are left with the following equality
$$\hskip-5cm - \mu_{d - 1}\varrho_{1}(X^{\ast}) - \mu_{d - 2}\varrho_{2}(X^{\ast}) - ... - \mu_{1}\varrho_{d - 1}(X^{\ast}) = \mu_{d}.$$
From the last equation we have the following equality
$$\hskip-3cm \mu_{1}\varrho_{d - 1}(X^{\ast}) = - \mu_{2}\varrho_{d - 2}(X^{\ast}) - ... - \mu_{d - 2}\varrho_{2}(X^{\ast}) - \mu_{d - 1}\varrho_{1}(X^{\ast}) - \mu_{d}.$$

\hskip-0.6cm Inserting this equality into the second equation of the system (3.4) we have
$$\mu_{d + 1} = \mu_{d}(x_{d - 1} - \varrho_{1}(X^{\ast})) + \mu_{d - 1}(\varrho_{1}(X^{\ast})x_{d - 1} - \varrho_{2}(X^{\ast})) + ... + \mu_{2}(\varrho_{d - 2}(X^{\ast})x_{d - 1} - \varrho_{d - 1}(X^{\ast}))$$ $$ \hskip-3cm + \left(-\mu_{2}\varrho_{d - 2}(X^{\ast}) - ... - \mu_{d - 2}\varrho_{2}(X^{\ast}) - \mu_{d - 1}\varrho_{1}(X^{\ast}) - \mu_{d}\right)x_{d - 1}.$$
Again, observe that all the terms which contain a product of a symmetric polynomial with the node $x_{d - 1}$ cancel each other. Hence we are left with the following equality
$$\hskip-5cm - \mu_{d}\varrho_{1}(X^{\ast}) - \mu_{d - 1}\varrho_{2}(X^{\ast}) - ... - \mu_{2}\varrho_{d - 1}(X^{\ast}) = \mu_{d + 1}.$$

Continuing in this way we can extract from the system of equations (3.4) the following system
\vskip-0.2cm
$$\hskip-3.15cm - \mu_{d + k - 2}\varrho_{1}(X^{\ast}) - \mu_{d + k - 3}\varrho_{2}(X^{\ast}) - ... - \mu_{k}\varrho_{d - 1}(X^{\ast}) = \mu_{d + k - 1}$$
where $k = 0,...,d - 1$. The last system of equations can be written in the following matrix form
$$\left(\begin{array}{ccccc}
\mu_{0} & \mu_{1} & ... & \mu_{d - 2} & \mu_{d - 1}\\
\mu_{1} & \mu_{2} & ... & \mu_{d - 1} & \mu_{d}\\
& .........\\
\mu_{d - 1} & \mu_{d} & ... & \mu_{2d - 3} & \mu_{2d - 2}\\
\end{array}\right)
\left(\begin{array}{c}
\varrho_{d - 1}(X^{\ast})\\
\varrho_{d - 2}(X^{\ast})\\
........\\
\varrho_{1}(X^{\ast})\\
1
\end{array}\right)
 = \left(\begin{array}{c}
0 \\
0\\
...\\
0\\
0
\end{array}\right).$$
Since, by our main assumption, the matrix in the left hand side of the last equation is non degenerate and hence we obviously arrive to a contradiction. Thus, if a point $X^{\ast} = (x_{1},x_{2},...,x_{d - 1})$ in $\Bbb R^{d - 1}$ satisfies the system of equations (3.3) then $P(X^{\ast})\neq0$ and hence $a_{d}$ tends to infinity. This proves Theorem 2.1. \hskip5cm $\square$

\vskip0.5cm

\textbf{Proof of Theorem 2.2:} From Proposition 1.2 we know that the symmetric polynomials $\sigma_{1},...,\sigma_{d}$ can be parameterized on $S_{2d - 2}^{X}$ as follows
$$\hskip-5cm\sigma_{d - k + 1} = \frac{(- 1)^{d + k}\mathcal{M}_{d,k}}{\det\mathcal{M}}t + b_{d - k + 1}, t\in A_{\mu}, k = 1,...,d$$
where $b_{k}, k = 1,...,d$ are constants which depend only on $\mu$ and are independent of $t$ (see formula (1.9)) and where $\mathcal{M}_{i,j}$ denotes the minor of the entry in the $i$-th row and $j$-th column of the matrix $\mathcal{M}$. Since, by assumption, $\mathcal{M}_{d,d}\neq 0$, Theorem 2.2 is a consequence of the following proposition.

\begin{prop}

Let $x_{1} = x_{1}(t),...,x_{n} = x_{n}(t), t\in\Bbb R$ be $n$ continuous functions which satisfy the following identities
\begin{equation}\sigma_{k}(X(t)) = ( - 1)^{k}\sum_{1\leq i_{1} < i_{2} < ... < i_{k}\leq n}x_{i_{1}}(t)x_{i_{2}}(t)...x_{i_{k}}(t) = a_{k}t + b_{k}, k = 1,...,n\end{equation}
where $a_{k},b_{k}\in\Bbb R$ and $a_{1} \neq 0$. Then, as $t\rightarrow\pm\infty$ there is at most one function $x_{i} = x_{i}(t)$ which tends to infinity.

\end{prop}

Before proving Proposition 3.1 we need the following lemma which is a direct consequence of the Budan-Fourier Theorem:

\begin{lem}

For any polynomial $P$ of degree $n$ and a point $x\in\Bbb R$, if $\nu_{P}(x)$ denotes the number of sign changes in the components of the vector $(P(x),P'(x),...,P^{(n)}(x))$ then the number of zeros of $P$ with multiplicity in the interval $(a,b]$ is less than or equal to $\nu_{P}(a) - \nu_{P}(b)$.

\end{lem}

\textbf{Proof of Proposition 3.1}: We can assume with out loss of generality that $a_{1} > 0$ since otherwise we can replace each $x_{i}$ with $-x_{i}$. Observe that from equation (3.5) it follows that for each $\lambda\in\Bbb R$ we have
\begin{equation}(\lambda - x_{1}(t))(\lambda - x_{2}(t))...(\lambda - x_{n}(t)) = \lambda^{n} + (a_{1}t + b_{1})\lambda^{n - 1} + ... + a_{n}t + b_{n}.\end{equation}
Hence, if we denote by $P_{t} = P_{t}(\lambda)$ the polynomial in the right hand side of equation (3.6) then we need to prove that as $t\rightarrow\infty$ (or $t\rightarrow-\infty$) all the roots of $P_{t}$ will be bounded except maybe only one root. Observe that by Lemma 3.2 the number of roots of the polynomial $P_{t}$ at the ray $(\lambda,\infty)$ is less than or equal to $\nu_{P_{t}}(\lambda) - \nu_{P_{t}}(\infty)$. Since obviously $\nu_{P_{t}}(\infty) = 0$ we only need to estimate $\nu_{P_{t}}(\lambda)$. Explicitly we have
$$\hskip-9cm(P_{t}(\lambda), P_{t}'(\lambda),..., P_{t}^{(n)}(\lambda))^{T}$$
$$ = \left(\begin{array}{c}\hskip-5.3cm\lambda^{n} + b_{1}\lambda^{n - 1} + ... + b_{n} + (a_{1}\lambda^{n - 1} + a_{2}\lambda^{n - 2} + ... + a_{n})t\\
n\lambda^{n - 1} + (n - 1)b_{1}\lambda^{n - 2} + ... + b_{n - 1} + \left[(n - 1)a_{1}\lambda^{n - 2} + (n - 2)a_{2}\lambda^{n - 3} + ... + a_{n - 1}\right]t\\
\hskip-11.7cm...........................\\
\hskip-10.3cm n!\lambda + (n - 1)!\cdot(a_{1}t + b_{1})\\
\hskip-14.475cm n!\end{array}\right)$$
\begin{equation} \hskip-9.7cm = \left(\begin{array}{c}
\hskip-1.6cm R(\lambda) + Q(\lambda)t\\
\hskip-1.5cm R'(\lambda) + Q'(\lambda)t\\
\hskip-2.6cm..........\\
R^{(n - 1)}(\lambda) + Q^{(n - 1)}(\lambda)t\\
\hskip-2.8cm R^{(n)}(\lambda)\\
\end{array}\right)\end{equation}
where
$$\hskip-1cm R(\lambda) = \lambda^{n} + b_{1}\lambda^{n - 1} + ... + b_{n}, Q(\lambda) = a_{1}\lambda^{n - 1} + a_{2}\lambda^{n - 2} + ... + a_{n}.$$
Observe that if for $\lambda\in\Bbb R$ we have $Q(\lambda) > 0,..., Q^{(n - 1)}(\lambda) > 0$ then obviously by taking $t$ large enough all the components of the vector defined by (3.7) will be positive and thus $\nu_{P_{t}}(\lambda) = 0$. Since $a_{1} > 0$ it follows that for each $k$ there exists $\lambda_{k}$ such that $Q^{(k)}(x) > 0$ for $x \geq \lambda_{k}$. Hence, by choosing $\lambda' = \max(\lambda_{0},...,\lambda_{n - 1})$ and $t$ large enough all the components of the vector (3.7) will be positive and thus $\nu_{P_{t}}(\lambda) = 0, \forall \lambda\geq\lambda'$ which will imply in particular that for $t$ large enough the polynomial $P_{t} = P_{t}(\lambda)$ does not have any roots for $\lambda\geq \lambda'$ where $\lambda'$ does not depend on $t$. Hence, at this point we proved that non of the roots $x_{i} = x_{i}(t)$ can tend to $+\infty$ as $t\rightarrow\infty$. We need to check how many roots can tend to $-\infty$ as $t\rightarrow\infty$.

For any $\lambda\in\Bbb R$ observe that by Lemma 3.2 the number of roots of the polynomial $P_{t}$ at the ray $(-\infty,\lambda)$ is less than or equal to $\nu_{P_{t}}(-\infty) - \nu_{P_{t}}(\lambda) = n - \nu_{P_{t}}(\lambda)$. Our aim now is to choose $\lambda$ such that the vector (3.7) will have exactly $n - 1$ sign changes (and thus $\nu_{P_{t}}(\lambda) = n - 1$). Observe that since $a_{1} > 0$ we can choose as before $\lambda$ in $\Bbb R$ which is small enough and which does not depend on $t$ such that
$$Q(\lambda) > 0, Q'(\lambda) < 0, Q''(\lambda) > 0,..., Q^{(n - 2)} < 0, Q^{(n - 1)}(\lambda) > 0$$
or
$$Q(\lambda) < 0, Q'(\lambda) > 0, Q''(\lambda) < 0,..., Q^{(n - 2)} < 0, Q^{(n - 1)}(\lambda) > 0$$
where the first case corresponds to the case where $n$ is odd and the second case corresponds to the case where $n$ is even. Thus, by taking $t$ large enough the signs of the vector (3.7) will have the form $(+-+-...+-++)$ or the form $(-+-+...-+-++)$. In either case the vector (3.7) will have exactly $n - 1$ sign changes and thus $\nu_{P_{t}}(\lambda) = n - 1$. Thus, from Lemma 3.2 it follows that the polynomial $P_{t}$ has at most $\nu_{P_{t}}(-\infty) - \nu_{P_{t}}(\lambda) = n - (n - 1) = 1$ roots at the ray $(-\infty,\lambda)$. This implies that only one of the functions $x_{i} = x_{i}(t)$ can tend to $-\infty$ as $t\rightarrow\infty$. Hence we proved Proposition 3.1 for the case where $t\rightarrow\infty$.

If $t\rightarrow-\infty$ we can just replace each function $x_{i} = x_{i}(t)$ with $y_{i} = y_{i}(t) = x_{i}(- t)$ and use the fact that the functions $y_{i}, i = 1,...,n$ satisfy a similar set of equations as (3.5). $\hskip13.8cm\square$

\vskip0.5cm

It can be proved from the Budan-Fourier Theorem that in fact the number of roots of a polynomial $P$ at the interval $(a,b]$ is equal to $\nu_{P}(a) - \nu_{P}(b) - 2k$ where $k$ is a nonnegative integer. Hence, from the proof of Proposition 3.1 we have the following corollary.

\begin{cor}

Let $x_{1} = x_{1}(t),...,x_{n} = x_{n}(t), t\in\Bbb R$ be $n$ continuous functions which satisfy the set of identities (3.5) where $a_{k},b_{k}\in\Bbb R$ and $a_{1} \neq 0$. Then, as $t\rightarrow\infty$ (or $t\rightarrow-\infty$) there is exactly one function $x_{i} = x_{i}(t)$ which tends to infinity.

\end{cor}

\section{The Special Cases $d = 2, 3$}

\hskip0.6cm For the special cases where $d = 2, 3$ we would like to analyse the projections $S_{2d - 2}^{X}$, of the Prony curves $S_{2d - 2}$, to the nodes parameter space $\mathcal{P}_{d}^{X}$ and answer the following two questions:
\vskip0.3cm
A. For which points $\mu\in\Bbb R^{2d - 1}$ collision of nodes on $S_{2d - 2}^{X}(\mu)$ actually occurs?
\vskip0.2cm
B. For which points $\mu\in\Bbb R^{2d - 1}$ the projections $S_{2d - 2}^{X}(\mu)$ are bounded?
\vskip0.2cm
\hskip-0.6cm We also give some examples to illustrate the main results obtained in Sect. 3. We will always assume that the corresponding Hankel matrix $\mathcal{M}$ for the vector $\mu$ is non degenerate.\\

\textbf{The case} $d = 2$: For $d = 2$, $S_{2}(\mu)$ is given as the set of solutions to the following system of equations

$$\hskip-10.75cm a_{1} + a_{2} = \mu_{0},$$
\vskip-0.6cm
$$\hskip-10.0cm a_{1}x_{1} + a_{2}x_{2} = \mu_{1},$$
$$\hskip-10.1cm a_{1}x_{1}^{2} + a_{2}x_{2}^{2} = \mu_{2}$$
and the determinant of its corresponding Hankel matrix $\mathcal{M}$ is given by
\vskip-0.2cm
$$\hskip-7.3cm \det \mathcal{M} =
\left|\begin{array}{cc}
\mu_{0} & \mu_{1}\\
\mu_{1} & \mu_{2}
\end{array}\right| = \mu_{0}\mu_{2} - \mu_{1}^{2}.$$
By Lemma 1.1 the projection $S_{2}^{X}(\mu)$ of $S_{2}(\mu)$ to the nodes parameter space $\mathcal{P}_{2}^{X}(\mu)$ is given by
\vskip-0.2cm
\begin{equation}\hskip-7.4cm\mu_{0}x_{1}x_{2} - \mu_{1}(x_{1} + x_{2}) + \mu_{2} = 0.\end{equation}
In order to determine for which points $\mu = (\mu_{0}, \mu_{1}, \mu_{2})$ there is a collision of nodes observe that in terms of the symmetric polynomials $\sigma_{1}$ and $\sigma_{2}$ the nodes $x_{1}$ and $x_{2}$ collide if and only if the polynomial
\vskip-0.2cm
$$\hskip-5.9cm Q(z) = (z - x_{1})(z - x_{2}) = z^{2} + \sigma_{1}z + \sigma_{2}$$
has a double root which occurs if and only if its discriminant $\triangle(\sigma_{1}, \sigma_{2}) = \sigma_{1}^{2} - 4\sigma_{2}$ vanishes. Hence, in the terms of the symmetric polynomials it follows from equation (4.1) that a collusion of nodes occurs if and only if the line
\begin{equation}\hskip-9.3cm\mu_{0}\sigma_{2} + \mu_{1}\sigma_{1} + \mu_{2} = 0\end{equation}
intersects the parabola $\triangle(\sigma_{1}, \sigma_{2}) = 0$. An intersection occurs if and only if the following equation
\vskip-0.2cm
$$\hskip-9cm\mu_{0}\sigma_{1}^{2} + 4\mu_{1}\sigma_{1} + 4\mu_{2} = 0$$
has at least one real root which, in case where $\det \mathcal{M}\neq0$, occurs if and only if $\det \mathcal{M} = \mu_{0}\mu_{2} - \mu_{1}^{2} < 0$. This answers Question A for $d = 2$.

In order to determine for which points $\mu = (\mu_{0}, \mu_{1}, \mu_{2})$ the projection $S_{2}^{X}(\mu)$ is bounded observe that in terms of the symmetric polynomials $\sigma_{1}$ and $\sigma_{2}$ each point $x$ in $S_{2}^{X}(\mu)$ corresponds, by the Vieta mapping, to a point $(\sigma_{1}, \sigma_{2})$ on the line (4.2) for which $\triangle(\sigma_{1}, \sigma_{2}) > 0$. Since the intersection of the line (4.2) and the set $\triangle(\sigma_{1}, \sigma_{2}) > 0$ is never bounded (unless $\mu_{0} = \mu_{1} = 0, \mu_{2}\neq0$ for which $\det\mathcal{M} = 0$) and the Vieta mapping maps bounded sets into bounded sets it follows that the inverse image of this intersection, i.e. the set $S_{2}^{X}(\mu)$, is also unbounded. Hence, there are no points $\mu\in\Bbb R^{3}$ for which $S_{2}^{X}(\mu)$ is bounded in case where $\det\mathcal{M}\neq0$. This answers Question B for the case $d = 2$.
\vskip0.3cm
\begin{example}

Let us illustrate Theorem 2.1 for the case $d = 2$. For $\mu = (0, 1, 0)$ the Prony curve $S_{2}(\mu)$ is given by the following system of equations
\vskip-0.2cm
$$\hskip-11.3cm a_{1} + a_{2} = 0,$$
$$\hskip-10.5cm a_{1}x_{1} + a_{2}x_{2} = 1,$$
$$\hskip-10.5cm a_{1}x_{1}^{2} + a_{2}x_{2}^{2} = 0.$$
Observe that
\vskip-0.2cm
$$\hskip-8.5cm \left|\begin{array}{cc}
\mu_{0} & \mu_{1}\\
\mu_{1} & \mu_{2}
\end{array}\right|
 = \left|\begin{array}{cc}
0 & 1\\
1 & 0
\end{array}\right|\neq 0.$$

Hence, one can expect that if the two nodes $x_{1}$ and $x_{2}$ collide on $S_{2}(\mu)$ then the amplitudes $a_{1}$ and $a_{2}$ will tend to infinity. By using some elementary algebraic manipulations one can show that $S_{2}(\mu)$ has the following parametrization

$$\hskip-6.5cm S_{2}(\mu) = \left\{\left(- \frac{1}{2t}, \frac{1}{2t}, - t, t\right): t > 0\right\}.$$
Observe that the nodes $x_{1}$ and $x_{2}$ collide as $t\rightarrow0^{+}$ in which case the amplitudes indeed tend to infinity.

\end{example}

\begin{example} Figure 1 below illustrates the projections $S_{2}^{X}(\mu)$ (drawn in red) to $\mathcal{P}_{d}^{X}$ of the Prony curves $S_{2}(\mu)$ for four different points $\mu = (\mu_{0}, \mu_{1}, \mu_{2})\in\Bbb R^{3}$ (we draw in blue the projections to the nodes parameter space $\{(x_{1}, x_{2}): x_{2} < x_{1}\}$). The images of these projections under the Vieta mapping $\mathcal{V}_{d}$ are the lines given in the upper part of Figure 1 and, as we have just proved, each intersection of a line $l$ with the parabola $\triangle(\sigma_{1}, \sigma_{2}) = 0$ corresponds to a collision of nodes in the preimage $\mathcal{V}_{d}^{-1}(l)$ as Figure 1 shows.

Observe also that, as accordance with Theorem 2.2, on the curves which are the connected components of any projection $S_{2}^{X}(\mu)$ to the nodes parameter space $\mathcal{P}_{d}^{X}$ only one node tends to infinity at a time except from the last case $x_{1} + x_{2} - 1 = 0$ where $\mu_{0} = 0$ for which the sufficient condition of Theorem 2.2 is not satisfied. In fact, the only cases for which the nodes $x_{1}$ and $x_{2}$ on $S_{2}^{X}(\mu)$ tend simultaneously to $\pm\infty$ occur when $\mu_{0} = 0$ or equivalently when the image of a projection $S_{2}^{X}(\mu)$, under the Vieta mapping $\mathcal{V}_{d}$, is contained in a vertical line. \end{example}

\textbf{The case} $d = 3$: For $d = 3$, $S_{4}(\mu)$ is given as the set of solutions to the following system of equations
\vskip-0.2cm
$$\hskip-9.7cm a_{1} + a_{2} + a_{3} = \mu_{0},$$
$$\hskip-8.5cm a_{1}x_{1} + a_{2}x_{2} + a_{3}x_{3} = \mu_{1},$$
$$\hskip-8.5cm a_{1}x_{1}^{2} + a_{2}x_{2}^{2} + a_{3}x_{3}^{2} = \mu_{2},$$
$$\hskip-8.5cm a_{1}x_{1}^{3} + a_{2}x_{2}^{3} + a_{3}x_{3}^{3} = \mu_{3},$$
$$\hskip-8.6cm a_{1}x_{1}^{4} + a_{2}x_{2}^{4} + a_{3}x_{3}^{4} = \mu_{4}$$

and by Lemma 1.1 the projection $S_{4}^{X}(\mu)$ of $S_{4}(\mu)$ to the nodes parameter space $\mathcal{P}_{3}^{X}(\mu)$ is given by
$$\hskip-1.5cm\mu_{0}x_{1}x_{2}x_{3} - \mu_{1}(x_{1}x_{2} + x_{1}x_{3} + x_{2}x_{3}) + \mu_{2}(x_{1} + x_{2} + x_{3}) - \mu_{3} = 0,$$
\vskip-0.8cm
\begin{equation} \end{equation}
\vskip-0.8cm
$$\hskip-1.5cm\mu_{1}x_{1}x_{2}x_{3} - \mu_{2}(x_{1}x_{2} + x_{1}x_{3} + x_{2}x_{3}) + \mu_{3}(x_{1} + x_{2} + x_{3}) - \mu_{4} = 0.$$

\begin{figure}
	\centering
	\includegraphics[scale=0.8]{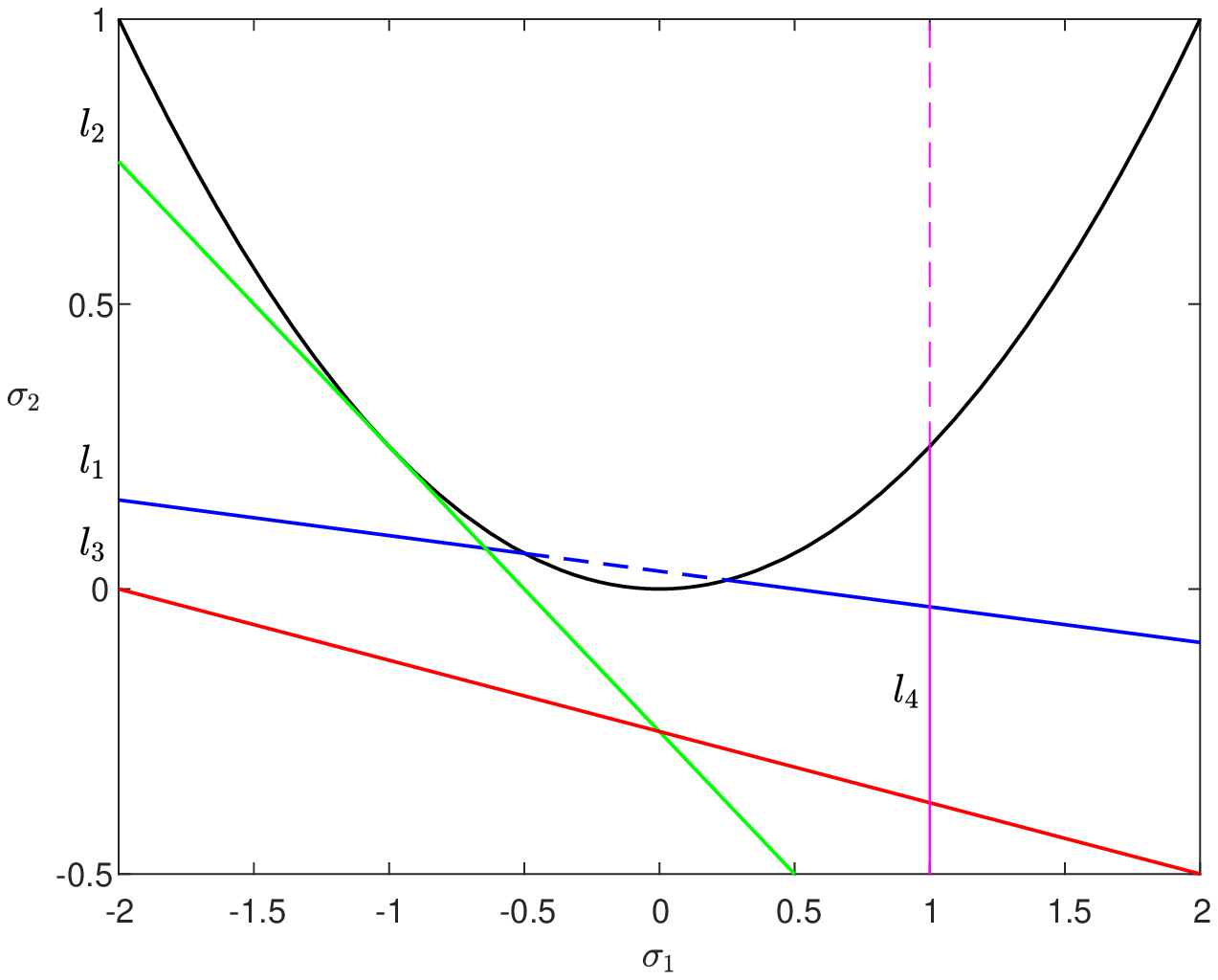}
	\includegraphics[scale=0.8]{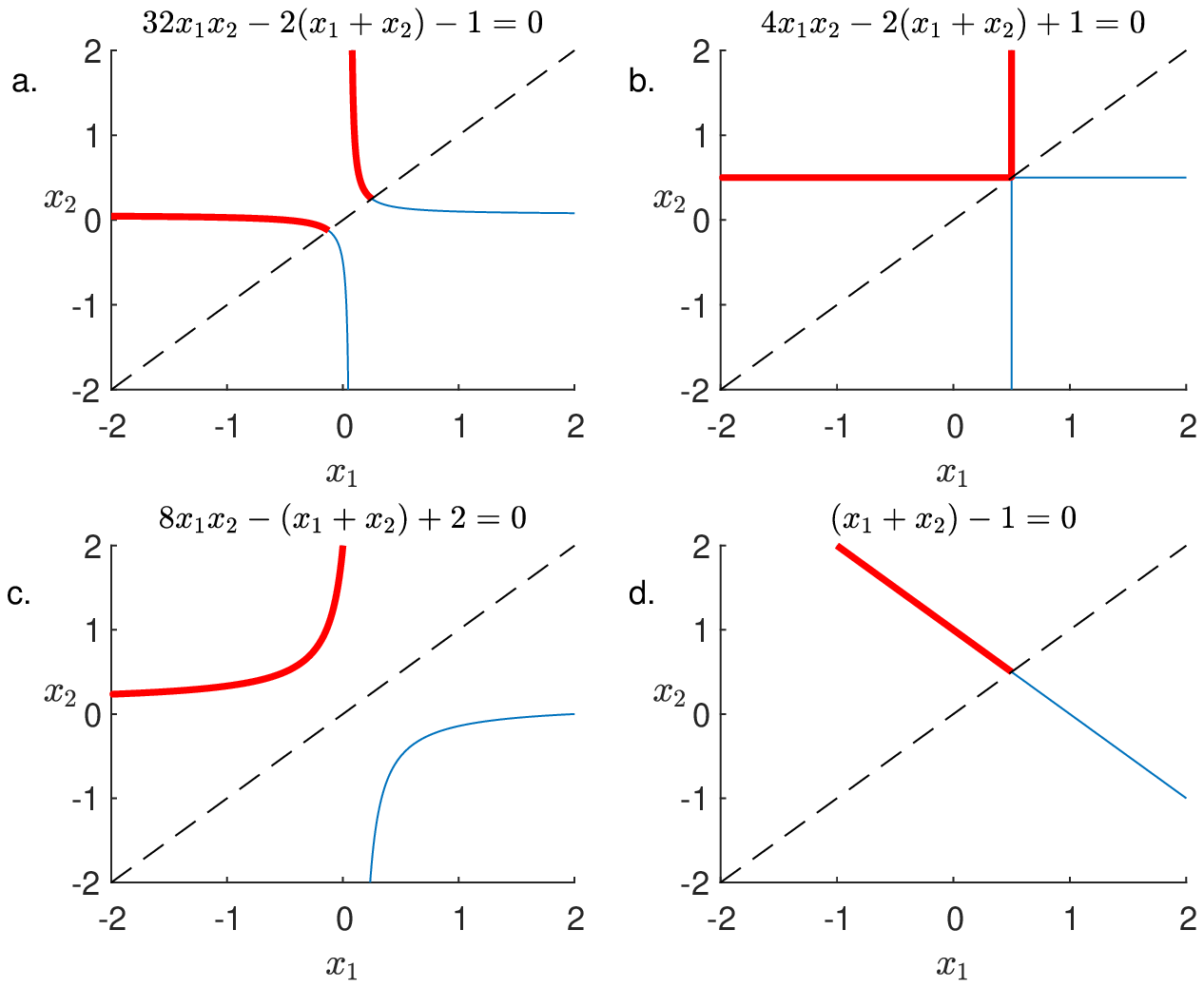}
	\caption{Typical projections of Prony curves.} 
	\label{fig.iso}
\end{figure}

In order to determine for which points $\mu = (\mu_{0}, \mu_{1}, \mu_{2}, \mu_{3}, \mu_{4})$ there is a collision of nodes observe that in terms of the symmetric polynomials $\sigma_{1}, \sigma_{2}$ and $\sigma_{3}$ the nodes $x_{1}$ and $x_{2}$ collide if and only if the polynomial
$$\hskip-3cm Q(z) = (z - x_{1})(z - x_{2})(z - x_{3}) = z^{3} + \sigma_{1}z^{2} + \sigma_{2}z + \sigma_{3}$$
has a double real root which occurs if and only if its discriminant
$$\hskip-3.3cm \triangle(\sigma_{1}, \sigma_{2}, \sigma_{3}) = 27\sigma_{3}^{2} + 4\sigma_{2}^{3} - \sigma_{1}^{2}\sigma_{2}^{2} + 4\sigma_{1}^{3}\sigma_{3} - 18\sigma_{1}\sigma_{2}\sigma_{3}$$
vanishes. In terms of the symmetric polynomials $\sigma_{1}, \sigma_{2}, \sigma_{3}$ it follows, from equation (4.3), that a collision of nodes occurs if and only if the line
$$\hskip-8cm \mu_{0}\sigma_{3} + \mu_{1}\sigma_{2} + \mu_{2}\sigma_{1} = - \mu_{3},$$
\vskip-0.8cm
\begin{equation} \end{equation}
\vskip-0.8cm
$$\hskip-8cm \mu_{1}\sigma_{3} + \mu_{2}\sigma_{2} + \mu_{3}\sigma_{1} = - \mu_{4}$$
intersects the surface $\triangle(\sigma_{1}, \sigma_{2}, \sigma_{3}) = 0$. Using Proposition 1.2 for the special case where $d = 3$ it follows that the line (4.4) has the following parametrization
$$\sigma_{1} = \frac{1}{|\mathcal{M}|}
\left|\begin{array}{ccc}
\mu_{0} & \mu_{1} & \mu_{3}\\
\mu_{1} & \mu_{2} & \mu_{4}\\
\mu_{2} & \mu_{3} & - t
\end{array}\right|,
\sigma_{2} = \frac{1}{|\mathcal{M}|}
\left|\begin{array}{ccc}
\mu_{0} & \mu_{3} & \mu_{2}\\
\mu_{1} & \mu_{4} & \mu_{3}\\
\mu_{2} & - t & \mu_{4}
\end{array}\right|,
\sigma_{3} = \frac{1}{|\mathcal{M}|}
\left|\begin{array}{ccc}
\mu_{3} & \mu_{1} & \mu_{2}\\
\mu_{4} & \mu_{2} & \mu_{3}\\
- t & \mu_{3} & \mu_{4}
\end{array}\right|$$
where $t\in A_{\mu}$. Inserting the above expressions for the variables $\sigma_{1}, \sigma_{2}$ and $\sigma_{3}$ into the equation $\triangle(\sigma_{1}, \sigma_{2}, \sigma_{3}) = 0$ we obtain an equation of the form
\begin{equation}\hskip-1.5cm P_{\mu}(t) = P_{8}(\mu)t^{4} + P_{9}(\mu)t^{3} + P_{10}(\mu)t^{2} + P_{11}(\mu)t + P_{12}(\mu) = 0\end{equation}
where $P_{k}$ is a homogenous polynomial in the variable $\mu$ of degree $k$. Since equation (4.5) is the restriction of the discriminant $\triangle(\sigma_{1}, \sigma_{2}, \sigma_{3})$ to the line (4.4) it follows that for every root $t_{0}$ to equation (4.5) at least two nodes from $x_{1}(t_{0}), x_{2}(t_{0})$ or $x_{3}(t_{0})$ coincide and this in particular implies that all these nodes are real at $t = t_{0}$. Since the set $A_{\mu}$ consists of all $t\in\Bbb R$ such that all the nodes $x_{1},x_{2}$ and $x_{3}$ are real and distinct it follows that $t_{0}$ is a boundary point of $A_{\mu}$. Thus, the condition that $t\in A_{\mu}$ does not restrict the inclusion of any of the roots of equation (4.5) when considering nodes collision.

Thus, a collision of nodes occurs if and only if the polynomial $P_{\mu}$ in the left hand side of equation (4.5) has at least one real root. This answers Question A for $d = 3$.

Contrary to the case $d = 2$, for the case $d = 3$ we preferred not to give an explicit relation on the parameters $\mu_{0},...,\mu_{4}$ for determining when collision of nodes occurs (or equivalently when $P_{\mu}$ has at least one real root) since this relation turns out to be overly complicated. One would hope that this relation could be written in a more compact form by expressing it in terms of the determinant of the matrix $\mathcal{M}$ and its minors.

Our aim now is to determine for which points $\mu = (\mu_{0}, \mu_{1}, \mu_{2}, \mu_{3}, \mu_{4})$ the projection $S_{4}^{X}(\mu)$ is bounded. Observe that in terms of the symmetric polynomials $\sigma_{1}, \sigma_{2}$ and $\sigma_{3}$ each point $x$ in $S_{4}^{X}(\mu)$ corresponds, by the Vieta mapping, to a point $(\sigma_{1}, \sigma_{2}, \sigma_{3})$ on the line (4.4) for which $\triangle(\sigma_{1}, \sigma_{2}, \sigma_{3}) < 0$ (the last condition on the discriminant guarantees that all the roots of $Q$ all real and distinct). By a direct computation we obtain that
$$\hskip-2.3cm P_{8}(\mu) = 4\left|\begin{array}{cc}\mu_{0} & \mu_{1}\\ \mu_{1} & \mu_{2} \end{array}\right|^{3}\left|\begin{array}{cc}\mu_{1} & \mu_{2}\\ \mu_{2} & \mu_{3} \end{array}\right| - \left|\begin{array}{cc}\mu_{0} & \mu_{1}\\ \mu_{1} & \mu_{2} \end{array}\right|^{2}\left|\begin{array}{cc}\mu_{0} & \mu_{2}\\ \mu_{1} & \mu_{3} \end{array}\right|^{2}$$
$$\hskip-3.3cm = - \frac{\mu_{0}^{4}}{27}\left|\begin{array}{cc}\mu_{0} & \mu_{1}\\ \mu_{1} & \mu_{2} \end{array}\right|^{2}\triangle\left(\frac{3\mu_{1}}{\mu_{0}}, \frac{3\mu_{2}}{\mu_{0}}, \frac{\mu_{3}}{\mu_{0}}\right).$$
Hence, if
\vskip-0.2cm
\begin{equation}\hskip-8.3cm \mu_{0}\neq0, \left|\begin{array}{cc} \mu_{0} & \mu_{1} \\ \mu_{1} & \mu_{2}\end{array}\right|\neq0\end{equation}
then $S_{4}^{X}(\mu)$ is bounded in case where
\vskip-0.2cm
\begin{equation}\hskip-5cm K(\mu_{0}, \mu_{1}, \mu_{2}, \mu_{3}) := \triangle\left(\frac{3\mu_{1}}{\mu_{0}}, \frac{3\mu_{2}}{\mu_{0}}, \frac{\mu_{3}}{\mu_{0}}\right) < 0\end{equation}
and unbounded if $K(\mu_{0}, \mu_{1}, \mu_{2}, \mu_{3}) > 0$. Indeed, if (4.7) holds then the polynomial $P_{\mu}$ will be nonpositive only inside a bounded interval $I$ in the variable $t$. The polynomial $P_{\mu}$ is the discriminant of the polynomial $Q$ restricted to the line defined by the system of equations (4.4). Hence, $\triangle(\sigma_{1}(t), \sigma_{2}(t), \sigma_{3}(t)) \leq 0$ only for $t\in I$ and since all the roots of $Q$ are real if and only if $\triangle(\sigma_{1}, \sigma_{2}, \sigma_{3})\leq0$ it follows that our parametrization, in the variable $t$, for the variables $\sigma_{1}, \sigma_{2}$ and $\sigma_{3}$ will be inside the interval $I$. That is, the set $A_{\mu}$ which parameterizes the symmetric polynomials $\sigma_{1}, \sigma_{2}$ and $\sigma_{3}$ satisfies $A_{\mu}\subseteq I$ where $I$ is bounded. This implies in particular that $\sigma_{1}, \sigma_{2}$ and $\sigma_{3}$ are bounded and thus, from the definition of the Vieta mapping, the nodes $x_{1}, x_{2}$ and $x_{3}$ in the projection $S_{4}^{X}(\mu)$ will also be bounded.

In the exact same way we can show that if $K(\mu_{0}, \mu_{1}, \mu_{2}, \mu_{3}) > 0$ then $S_{4}^{X}(\mu)$ is unbounded. This answers Question B for the case $d = 3$ in the general case when $P_{8}(\mu)\neq0$. In the extreme case where $P_{8}(\mu) = 0$ then we can make the exact same analysis on the coefficients $P_{9},...,P_{12}$ that are left in order to determine whether $S_{4}^{X}(\mu)$ is bounded or not. The details are left for the reader.

\section{Appendix}

\textbf{Proof of Lemma 1.1}: For a fixed vector $\mu = (\mu_{0},...,\mu_{q})\in\Bbb R^{q + 1}$ our aim is to show that the projection $X$ of each point $(A, X)\in\mathcal{P}_{d}$, satisfying the system (1.3), to the nodes parameter space $\mathcal{P}_{d}^{X}$ satisfies the system (1.5) and vice versa that each point $X\in\mathcal{P}_{d}^{X}\subset\Bbb R^{d}$, satisfying the system (1.5), also satisfies the system (1.3) for some point $A = (a_{1},..., a_{d})\in \mathcal{P}_{d}^{A}$.

For the proof of the first direction let us take the projection $X$ to the nodes parameter space $\mathcal{P}_{d}^{X}$ of a point $(A, X)\in\mathcal{P}_{d}$ which satisfies the system (1.3). For each $z\in\Bbb C$ the symmetric polynomials $\sigma_{1},...,\sigma_{d}$ satisfy
\vskip-0.2cm
\begin{equation}\hskip-2cm Q(z) = (z - x_{1})(z - x_{2})...(z - x_{d}) = z^{d} + \sigma_{1}(X)z^{d - 1} + ... + \sigma_{d}(X).\end{equation}
Hence, it follows that for $d\leq l\leq q$ we have
\vskip-0.2cm
$$\hskip-3.5cm\sum_{i = 0}^{d}\mu_{l - i}\sigma_{i}(X) = \sum_{i = 0}^{d}\sigma_{i}(X)\sum_{j = 1}^{d}a_{j}x_{j}^{l - i} = \sum_{j = 1}^{d}a_{j}\sum_{i = 0}^{d}\sigma_{i}(X)x_{j}^{l - i}$$
$$\hskip-4.8cm = \sum_{j = 1}^{d}a_{j}x_{j}^{l - d}\sum_{i = 0}^{d}\sigma_{i}(X)x_{j}^{d - i} = \sum_{j = 1}^{d}a_{j}x_{j}^{l - d}Q(x_{j}) = 0$$
where in the notation $\sigma_{0}$ we mean that $\sigma_{0} = 1$ and where in the last passage we used the fact that the polynomial $Q$ vanishes at the points $x_{i}, i = 1,...,d$ as can be seen from equation (5.1). This proves the first direction in the proof of Lemma 1.1.

For the proof of the opposite direction, assume that $X\in\mathcal{P}_{d}^{X}$ satisfies equation (1.5), then we need to find a point $(A, Y)\in\mathcal{P}_{d}$ which satisfies the system (1.3) and such that its projection to the nodes parameter space $\mathcal{P}_{d}^{X}$ coincides with $X$, i.e., $Y = X$. Observe that by our assumption that $q\geq d$ it follows that for every point $(A, Y)\in\mathcal{P}_{d}$ which satisfies the system (1.3) the point $A$ is uniquely determined by the first $d$ equations of (1.3) which is a nondegenerate system of equations of Vandermonde's type since $Y\in\mathcal{P}_{d}^{X}$. Hence, we only need to show that by choosing $Y = X$ the last $q - d + 1$ equations of the system (1.3) are satisfied assuming that the first $d$ equations of (1.3) are satisfied and that $X$ is a solution to (1.5). First we will prove that if the first $d + k - 1$ equations of the system (1.3) are satisfied where $1\leq k\leq q - d + 1$ and $X$ satisfies the system (1.5) then the $(d + k)^{th}$ equation of the system (1.3) is satisfied. Since the first $d + k - 1$ equations in the system (1.3) are satisfied it follows in particular that the quantities $\mu_{k - 1},...,\mu_{k + d - 2}$ can be expressed by the moments of the point $X$. Thus, from the $k^{th}$ equation of the system (1.5) we have
$$\hskip-3cm\mu_{d + k - 1} = - \sum_{i = 1}^{d}\mu_{k + d - i - 1}\varrho_{i}(X) = - \sum_{i = 1}^{d}\sum_{j = 1}^{d}a_{j}x_{j}^{k + d - i - 1}\varrho_{i}(X)$$ $$\hskip-0.5cm = - \sum_{j = 1}^{d}\sum_{i = 1}^{d}a_{j}x_{j}^{k + d - i - 1}\varrho_{i}(X) = - \sum_{j = 1}^{d}a_{j}x_{j}^{k - 1}\left(Q(x_{j}) - x_{j}^{d}\right) = \sum_{j = 1}^{d}a_{j}x_{j}^{d + k - 1}$$
where in the fourth and fifth passages we used equation (5.1). Hence, we proved that the $(d + k)^{th}$ equation of the system (1.3) is satisfied. Now the opposite direction follows easily by induction. Indeed, Since by our assumption the first $d$ equations in (1.3) are satisfied then it follows that the $(d + 1)^{th}$ equation is satisfied. Assume now that the first $d + m$ equations in the system (1.3) are satisfied where $0\leq m\leq q - d$, then in particular the last $d$ equations in this family of equations are satisfied and thus the $(d + m + 1)^{th}$ equation is satisfied. This finishes the proof by induction and thus the system of equations (1.3) is satisfied. Hence, Lemma 1.1 is proved.

{}

\end{document}